\documentclass[12pt,preprint]{aastex} 

\usepackage{epsfig}

\def\Msolar{\mathrm{M_\odot}}
\begin{document}

\title{X-ray Bursts in Neutron Star and Black Hole Binaries from USA and RXTE Data: Detections 
and 
Upper Limits}

\author{D. Tournear\altaffilmark{1},
E. Raffauf,
E. D. Bloom, 
W. Focke, 
B. Giebels\altaffilmark{2}, 
G. Godfrey, 
P. M. Saz Parkinson,
K. T. Reilly}

\affil{Stanford Linear Accelerator Center, Stanford University, 
Stanford, CA 94309}
\vspace{0.7mm}
\vspace{0.5mm}
\author{K. S. Wood, 
P. S. Ray,
M. T. Wolff,
R. M. Bandyopadhyay\altaffilmark{3}$^,$\altaffilmark{4},   
M. N. Lovellette}
\vspace{0.5mm}
\affil{E. O. Hulburt Center for Space Research, 
	Naval Research Laboratory,
	Washington, DC 20375}
\vspace{0.7mm}
\and
\author{Jeffrey D. Scargle}
\affil{NASA/Ames Research Center, Moffett Field, CA 94035}
\altaffiltext{1}{tournear@SLAC.stanford.edu}
\altaffiltext{2}{Current address: Laboratoire Leprince-Ringuet,
Ecole Polytechnique, Palaiseau F-91128}
\altaffiltext{3}{NRC/NRL Cooperative Research Associate}
\altaffiltext{4}{Current address: Department of Astrophysics, Oxford University}

\begin{abstract}
Narayan and Heyl (2002) have developed a theoretical framework to
convert suitable upper limits on type I X-ray bursts from accreting
black hole candidates (BHCs) into evidence for an event horizon.
However, no appropriate observational limit exists in the literature.
In this paper we survey 2101.2 ks of data from the Unconventional
Stellar Aspect (USA) X-ray timing experiment and 5142 ks of data from 
the Rossi X-ray Timing Explorer (RXTE) experiment to obtain a formal
constraint of this type.  1122 ks of neutron star data yield a
population averaged mean burst rate of $1.69 \times 10^{-5}$ bursts
s$^{-1}$ while 6081 ks of BHC data yield a 95\% confidence level upper
limit of $4.9 \times 10^{-7}$ bursts s$^{-1}$.  This is the first published 
limit of this type for Black Hole Candidates.\\

Applying the theoretical framework of Narayan and Heyl (2002) we 
calculate regions of unstable luminosity where the neutron stars 
are expected to burst and the BHCs would be expected to burst if they 
had a surface.  In this unstable luminosity region 464 ks of neutron star data 
yield an averaged mean burst rate of $4.1 \times 10^{-5}$ bursts s$^{-1}$ 
and 1512 ks of BHC data yield a 95\% confidence level upper
limit of $2.0 \times 10^{-6}$ bursts s$^{-1}$, and a limit of $> 10 \ \sigma$
that BHCs do not burst with a rate similar to the rate of neutron stars
in these unstable regions.  This gives further evidence that BHCs do not have
surfaces unless there is some new physics occurring on their surface.
\end{abstract} 

\section{Introduction}
\label{sec:intro}
	
In the nuclear burning model for X-ray bursts, a well-defined surface
must exist for a type I X-ray burst to occur.
\citet{nh02} (hereafter NH02) performed a stability analysis of
accumulating fuel on the surface of a generic compact object and
showed that if Black Hole Candidates (BHCs) had surfaces, they would be expected to exhibit
X-ray bursts.  For this reason the lack of type I X-ray bursts in
BHCs in stellar systems can be considered as evidence for an event
horizon.  

BHCs are typically distinguished from neutron stars using
one of three methods.  The primary method is direct dynamical
measurements of their companions yielding masses for the compact
objects that are $\gtrsim 3 \Msolar$, where a stable equation of state
for a neutron star does not exist.  
The second method is that if timing and spectral
analysis of the radiation from the accretion onto the compact object 
resembles other known BHCs the compact
object is usually classified as a BHC.  This typically means very
little power above $\sim$500 Hz in a Fourier power density spectrum
\citep{sr00} and an energy spectrum that cannot be
explained by a low temperature blackbody with an emission radius of
$\sim$10 km {\citep{rbb+00}}.  Thirdly, if an object
exhibits type I burst behavior it is assumed to be a neutron star as
this is strong evidence for the presence of a surface.  This is the 
case for the compact object in GS 1826-238, which was first classified 
as a BHC based on timing and spectral behavior.  It was only reclassified
as a neutron star when type I X-ray bursts were detected ~{\citep{ubc+99}}. 
 

Unstable thermonuclear burning on the surface of a weakly magnetized
neutron star is the generally accepted model for type I X-ray bursts.
Material accretes onto the surface until it reaches densities
and temperatures sufficient for nuclear ignition. For some accretion rates, 
the burning is unstable and propagates around the star consuming
all of the available fuel, resulting in an X-ray burst.  This type of
X-ray burst is characterized by a fast rise time of  a
few seconds and a decay time longer than a few seconds (up to several minutes).  See
\citet{sb03}, \citet{hv75},
\citet{lvt95}, \citet{p83}, and 
\citet{b00} for a more detailed description. The type I
X-ray bursts discussed in this paper are different from the X-ray
superbursts recently discovered \citep[e.g.][]{kiv+02}.
Type I X-ray bursts occur when helium on the surface of a neutron star
ignites via the triple-alpha process forming carbon and oxygen.  
These bursts last $\sim$ a few seconds -- 100s of seconds.  Superbursts likely 
occur due to nuclear burning
of carbon or by some other nuclear processes and last for
several hours (see \citealt{cb01} and \citealt{sb02} for a review).  

In this paper we present observations of neutron star LMXBs and BHCs
from the USA Experiment as observational support for the work of NH02.
We survey both neutron star and BHC systems for X-ray bursts and
calculate burst rates for neutron stars and upper limits for the burst
rate in BHCs, both overall and in the unstable burning regions
calculated by NH02.  This provides quantitative results which are
important in any attempt to demonstrate formally the existence of
event horizons using the method of NH02.  In addition, it is important that
any evidence in support or contradiction of the black hole model be
aired to watch for evidence the standard picture may not be quite right \citep{pee02,bg02}.   

\section{Observations and Data Analysis}

\subsection{The USA experiment}
\label{sec:USA} 

USA is an X-ray timing experiment built jointly by the Naval Research
Laboratory and the Stanford Linear Accelerator Center for the dual
purpose of conducting studies of variability in X-ray sources and
exploring applications of X-ray sensor technology (see
\citealt{rwf+01} for more details).  USA was launched on
1999 February 23 aboard the Advanced Research and Global Observation
Satellite (ARGOS), and took data until the mission ended in 2000 November.  The
primary observing targets were bright galactic X-ray binaries, with
the goal of obtaining large exposures on a small number of sources.

The detector consists of two proportional counters sensitive in the
range 1--15 keV with an effective area of about 1000 cm$^2$ each at 3
keV.  One proportional counter failed very early in the mission and for
the rest of the mission only one proportional counter was used.  
All the observations in this paper were made with one proportional counter only.
Collimators define the field of view (FOV), which is approximately
1.2$^\circ$ FWHM circular.  The Crab Nebula gives about 3500 cts s$^{-1}$ at
the center of the field of view in one counter. 
USA event timing is accurate to better than 32 $\mu$s.
In the case of the LMXB Scorpius X$-$1 the count rates are so high that USA 
will auto-shutdown if it is directly pointed at the source. 
Thus, we must offset point at Scorpius X-1 so as to sample its source
flux at about the 15\% response point of the collimator.

\subsection{RXTE Experiment}
	The RXTE experiment is an X-ray satellite composed of three instruments: the Proportional 
counter array (PCA), the high energy X-ray timing experiment (HEXTE), and the all sky monitor 
(ASM).  The PCA is effective over the range 2-60 keV with 18\% energy resolution at 6 keV.  
Data from the HEXTE instrument was not used in this paper.  The 
ASM has three detectors sensitive in the 2-10 keV range and rotates to take a $\sim 90$ second 
exposure on $\sim 75$ sources several times daily.  More 
detailed information on RXTE can be found in ~\citet{jsg+96}.  


\subsection {Data Analysis}
\label{sec:data}

All binary sources with more than 30 ks of exposure time with the USA
detector were scanned for bursts, including seven BHCs and seventeen
neutron stars.  The data analyzed were selected based on low
background counting rate and a small $\leq 0.5^\circ$ pointing offset.
The selection criteria result in less than 30 ks of analyzed data on
some sources.  The list of selected data analyzed in this paper is
shown in Table \ref{tab:sources}.  The columns are the source name,
total time of good data analyzed, distance and mass of the source,
classification of the source (BHC) Black Hole Candidate, Neutron Star
(NS) and if a subdivision of Z or Atoll is clear.  Additional
information on the sources can be found in the references given.  To supplement the 
USA BHC data, the RXTE public BHC data was analyzed for each USA BHC source 
analyzed in this paper, all public data for these sources as of April 2003, were analyzed.

\subsubsection{Source Categorization Criteria}
\label{sec:criteria}
Because we are testing for the presence of bursts on objects purported to be black holes the 
classification criteria used must be crafted to avoid selection effects.  Most importantly, 
the sources must not be classified based on the presence or absence of type I bursts.  We have 
thus categorized our sources into neutron stars or black holes as follows.

\paragraph{Black Hole Candidates}
Each of the BHCs, with the exception of 4U~1630$-$472, have a known mass function or
other mass estimate that give a lower mass limit of the compact
object that exceed current theoretical limits for neutron stars.  Since 4U 1630$-$472
does not have a mass limit requiring it to be a BHC, the data from this source
is not used in calculating any of the bursting rates or bursting rate limits
used later in this paper.  Cygnus X-3 also has a debateable compact object.  Cygnus X-3 has a 
mass function of $2.3 \Msolar$ \citep{sgs96} which does not rule out the possibility that the 
compact object is a neutron star.  Therefore, the data of Cygnus X-3 is not used in 
calculating any burst rates or limits quoted in this paper.  These two sources are included in 
Table ~\ref{tab:sources} for
completeness, to let the curious reader know that we did search the data of these sources
and no X-ray bursts were detected.
\paragraph{Neutron Stars}
The following neutron stars were classified as such based on the presence of kilohertz 
Quasi-Periodic Oscillations (QPOs): Cygnus~X-2, Aquila~X-1, EXO~0748$-$676, 
GX~$354-0$~(4U~1728-34), 4U~0614$+$09, GX~349+2, Scorpius~X-1, 4U~1735$-$445, X1636$-$536, 
GX~5-1, GX~340+0, and GX~17+2.  
The generally accepted view is that these kHz QPOs represent in some way an orbital frequency 
around the compact object.  The frequency of an orbit around a compact object is:
\begin{equation}
\nu_{orb} = \left(\frac{G M}{4\pi^2r^3_{orb}}\right)^{1/2} \approx 1200 \mathrm{\ Hz\ } 
\left(\frac{r_{orb}}{15\mathrm{\ km}}\right)^{-3/2} M^{1/2}_{1.4}
\end{equation}
where, $M_{1.4}$ is the mass of the compact object in units of $1.4 \Msolar$.
From general relativity, no stable orbital motion is possible within the Innermost Stable 
Circular Orbit (ISCO).  $R_{ISCO} = 6\mathrm{G}M/c^2 \approx 12.5 M_{1.4}$ km.  At this orbit, 
the the frequency is $\nu_{ISCO} \approx (1580/M_{1.4})$ Hz ~\citep{van00}.  From this, one 
can see that an object too massive to exist as a neutron star, $M \gtrsim 3 \Msolar$, can not 
demonstrate QPOs in the kHz range.  For a list of known sources demonstrating kilohertz QPOs 
and a review of millisecond oscillations in X-ray binaries see ~\citet{van00}.

  The Rapid Burster is classified as a neutron star based on the presence of excess power in 
the kHz range during it's type II X-ray bursts as reported in \citet{gfl+97}.  The presence of 
significant noise above 500 Hz in an X-ray Power Density Spectrum (PDS) gives evidence for a 
neutron star according to \citet{sr00}.  In addition, spectral work from Chandra data gives 
evidence of a radius $\sim 10$ km for this source by looking at the type II X-ray bursts 
~\citep{mrf+01}, giving further evidence that this source should be classified as a neutron 
star.  

  The sources XB 1254-690 and Circinus X-1 do not have strong enough evidence to support that 
they are neutron stars independent of the past observed Type I X-ray bursts.  Therefore, these 
sources will be included in Table ~\ref{tab:sources} for completeness, but will not be 
included in any bursting rate or bursting rate limit calculations.

	
\subsubsection{Detecting Bursts}
\begin{figure}
\begin{center}
\plotone{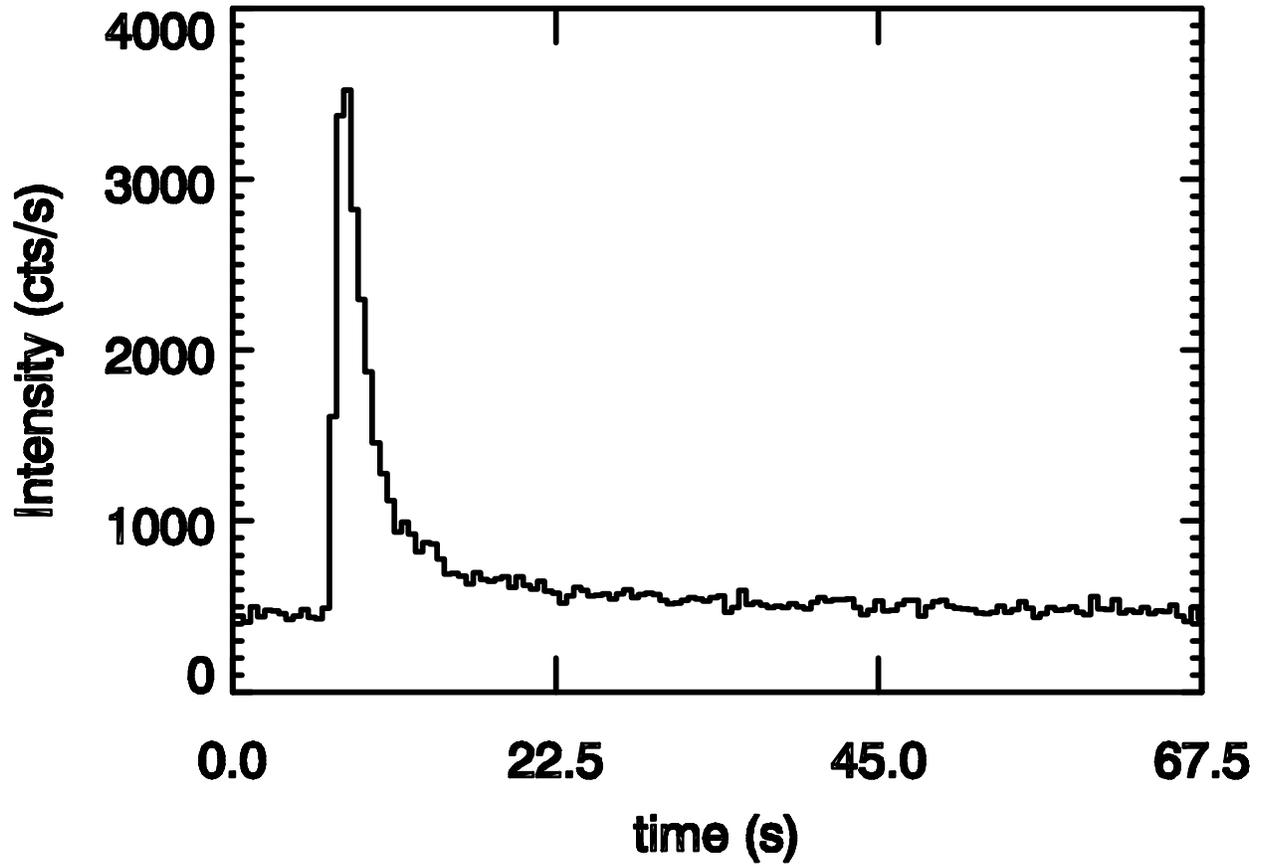}
\end{center}
\caption{
Example of a typical type I X-ray burst observed in 4U $1735-445$
by USA.}
\label{fig:RBburst}
\end{figure}

The data were searched for bursts by visually inspecting the light
curves, binned in 1 second intervals for USA data and $1/8$ second intervals for
RXTE data, of each observation several times.  The visual scan looked for
events that displayed properties consistent with type I X-ray bursts:
large changes in flux with Fast Rise Exponential Decay (FRED)
profiles, rise times of a few seconds, and decay times of several
seconds or more.  At the flux levels of all of the sources included in
this work, this method will easily detect all type I X-ray bursts.  Even if a burst in a BHC 
occurred with an order of magnitude less luminosity, as would be expected due to general 
relativistic effects see \citep{akl02}, we expect to have detected them visually.
Figure \ref{fig:RBburst} shows an example of a typical type I X-ray
burst found by USA in 4U $1735-445$.  We did not search the USA
data archive for superbursts.  The longest continuous observation by
USA was 20 minutes, with no less than 90 minutes between observations
of the same source, making a burst spanning several hours difficult to
detect.  Unless specified, the term ``X-ray burst'' used in this paper
refers to type I X-ray bursts.

\subsection{Observations of Neutron Star Systems}
\label{sec:ns}

Of the seventeen neutron stars that we studied (see Table
\ref{tab:sources}), seven were observed to have type I X-ray bursts.
These are: Aquila X-1, EXO 0748$-$676, GX 354$-$0,  the Rapid Burster,
4U 1735$-$445 , GX 3+1, and MXB 1659$-$298.  
The Rapid Burster and 
GX 354$-$0 are separated in the sky by only $0.54^\circ$. 
Thus, they will be in the USA FOV at the same 
time making it possible to mistake a burst in one  
source for a burst in the other source, albeit attenuated by 
a factor of 0.53. 
For this reason, the data for these two sources were double counted in our
analysis, meaning that an observation pointed at either source was considered to be
a simultaneous observation of both sources.  Therefore, the total observing time
of these two sources will be twice the sum of the respective pointed observations.
In theory, one may distinguish the Rapid Burster type I X-ray bursts from the GX~354$-$0
X-ray bursts.  The GX~354$-$0 bursts nearly always show radius expansion and are much
brighter than the Rapid Burster type I X-ray bursts~\citep{kuu03,fox03}.  In addition, the 
GX~354$-$0 bursts may display a harder spectrum than the Rapid Burster 
bursts~\citep{flr+01,fox03}.
However, we were not able to see definite evidence for radius expansion in the energy
spectra of any of these bursts, and we were not able to distinguish two groups of bursts
by any energy or timing analysis of the bursts from these sources, this may be due to the 
energy resolution of the USA experiment.  The intensity difference can not be used to 
distinguish these sources since a collimator effect would disguise the burst intensity.  
Therefore, since the Rapid Burster and GX~354$-$0 
are both known to display type I X-ray bursts regularly~\citep{lvt95,flr+01}, 
we can not distinguish the bursts from these two sources.  

 Several type I X-ray bursts
were observed in the vicinity of the Rapid Burster in addition to the
hundreds of type II bursts that occur continuously in this source when
it is in a bursting state.  We distinguished the type II bursts in this source
from the type I bursts by their regular, ``rapid-fire'', intervals.
Spectral differences can be used to give a more definite distinction between the 
types of X-ray bursts.  Type I X-ray bursts always have cooling blackbody spectra.
Type II X-ray bursts do not show softening throughout the burst characteristic of 
a cooling blackbody.  See
\citet{gfk+99,lvt93} for more information on the Rapid Burster and
type II X-ray bursts. Table \ref{tab:bursts} gives a list of bursts found in each source in 
our survey.

We did not detect bursts in ten of the seventeen neutron stars in our data:
Cygnus X-2, Circinus X-1, 4U 0614+09, GX 349+2, XB 1254$-$690, Scorpius
X-1, X1636$-$536,  GX 5$-$1, GX 340+0, GX 17+2.  Type I X-ray bursts
have been detected in the past for each of these neutron star systems
observed except for GX 349+2, Scorpius X-1, GX 5$-$1, and GX 340+0.
Using the data from all seventeen neutron stars, we calculate an overall 
neutron star bursting rate
$\lambda_{\mathrm{NS}} = 1.69\times10^{-5}$~bursts s$^{-1}$.  This results
in an observed average time between bursts in our data,
$R_{\mathrm{NS}}= 59.1$~ks, about 16.5~hours, from 1122~ks of data
(double counting the GX~354$-$0 and Rapid Burster data).

\subsection{Observations of Black Hole Candidates}
\label{sec:BHC}

Seven BHCs were searched for bursts: Cygnus X-1, XTE
J1118+480, GRS 1915+105, XTE J1859+226, XTE J1550$-$564, 4U 1630$-$472, and Cygnus
X-3.  The total USA observing time for all of these BHCs is 1022 ks, the total observing time 
for RXTE is 5477 ks.
No evidence of bursts was detected.  Data from 4U 1630$-$472 and Cygnus X-3 were not used in 
any burst limit calculations.  GRS 1915+105 showed several flares during 
the USA observations, close examination showed no distinct FRED profiles in these
flares that would signal a type I X-ray burst.  The USA observations were
scheduled for repeated sampling to obtain many short observations over
a long time period.  The RXTE Target of Opportunity (TOO) observations repeatedly sampled 
transient sources during outbursts.  As applied to transients this is particularly
effective for the sources studied here: bursting may not occur in all
ranges of mass accretion rate but the USA and RXTE observations sampled this
critical parameter over the maximum extent possible.  If BHCs burst in
a narrow range of mass accretion rate, we are likely to have sampled
that range.  The USA experiment was fortunate that over its lifetime
it was able to observe a number of very important transients,
including systems with the largest known physical dimensions (GRS 1915+105)
and the smallest (XTE J1118+480). USA and RXTE also devoted substantial time to
Cygnus X-1 and GRS 1915+105, so that there was ample opportunity for any rare modes of
bursting to be present in our observation of these sources.

Type I X-ray bursts are not a Poisson process.  The time between
bursts is a function of how long it takes to accumulate fuel to
reach a critical density and temperature where unstable nuclear
burning can occur.  Burst intervals have been observed to be regular and irregular,
and range from~$\sim 5$~min to days or longer~\citep{lvt95}, and are dependent on 
the accretion composition and rate \citep{sb02}.
However, since our observation times were short compared to typical
burst intervals and the revisit times were hours or longer, we are
unlikely to observe consecutive bursts.  Therefore, we treat these
observations using Poisson statistics in order to place a numerical
limit on the bursting rate. The probability of observing $n$ events of
a Poisson process is:
\begin{equation}
\label{eq:poisson}
	P(n) = \left(\lambda T\right)^{n} \frac{e^{-\lambda T}}{n!} 
\end{equation}
where: $\lambda$ is the rate of bursts (bursts s$^{-1}$) and $T$ is the
total observing time.  We will also define $R = 1/\lambda$ the average
time between bursts.  To set a limit on the burst rate to a
confidence level of $CL=95\%$ we calculate from eq. \ref{eq:poisson}
for $n=0$:
\begin{equation}
\label{eq:limit}
	\lambda = -\frac{\ln{(1-CL)}}{T}
\end{equation}
where $P = 1-CL$.  
For the values given, we calculate the upper limit of $\lambda_{\mathrm{BHC}}
= 4.9\times10^{-7}$ bursts s$^{-1}$ with a $95\%$ confidence level.  This is
the first published survey to place quantitative limits on the rate of
occurrence of bursts in a wide range of BHCs with a significant amount
of data.  As discussed in \citep{kni03} limits of these type emphasizing searches and null 
results are important to promulgate to the rest of the scientific community.

\section{Discussion}
\label{sec:discussion}

In this section we discuss these observations in the context of the
theoretical work of NH02.  We start by converting our results to a format
which allows easy comparison with theoretical predictions.  Next, we show
that our data in conjunction with the stability analysis of NH02, can
place a probablity-limit on the existence of a surface on a BHC.  
Finally, we check our neutron star results for consistency with the 
NH02 analysis.

\subsection{Luminosity Calculation}

From the 17 neutron stars analyzed we calculate a bursting rate of
$\lambda_{\mathrm{NS}} = 1.69\times10^{-5}$ bursts s$^{-1}$.  This is
comparable to rates found by other observers \citep[see][]{sb03}.  NH02 predict
that the occurrence of Type I X-ray bursts is a strong function of
$\log \left(\frac{L}{L_{\mathrm{Edd}}}\right)$, where $L_{\mathrm{Edd}}$ is the
Eddington luminosity of the source, $L_{\mathrm{Edd}} = 1.3\times
10^{38}M\mathrm{\ ergs~s}^{-1}$ (For H rich material) and $M$ is the mass of the compact
object in $\Msolar$. NH02 show that for certain
values of $\log \left(\frac{L}{L_{\mathrm{Edd}}}\right)$ there can be
stable burning and for other values there will be unstable burning
leading to type I X-ray bursts.  The exact ranges of $L$ for stable
accretion depend on the surface temperature of the compact object and
its radius.  For neutron stars, assuming a temperature at the base of
the accretion layer of $10^8$ K, the region of instability lies
between $-1.5 \lesssim \log \left(\frac{L}{L_{\mathrm{Edd}}}\right)
\lesssim -0.5$.

For each of the sources analyzed, we made a light curve in units of
$\log \left(\frac{L}{L_{\mathrm{Edd}}}\right)$ for each observation.
We then looked at the fraction of time sources were in the range of
$L$ where NH02 predict bursts should occur.
In order to calculate the luminosity of
the source, we fit a spectrum to a typical observation of the source
using XSPEC \citep{da01}.  If distinct states in the
source were evident from the source's hardness ratio, then
observations from these states were fit separately.  Most sources were
fit using the Bulk Motion Comptonization (BMC)
\citep{tmk97} model in XSPEC with absorption and iron
emission lines as needed.  The BMC was the best fitting model to most
of our data.  After fitting a model to the data we obtained the model
flux in the $0.2 - 30.0$ keV band.  Since we know the USA count rate
of the source for this flux, we are able to calculate a conversion
factor from USA rate to source flux.
\begin{equation}
	1\ \mathrm{USA\ cts\ s^{-1}} \approx C \times 10^{-12} \mathrm{ergs}\ \mathrm{cm^{-2}\  
s^{-1}} 
\end{equation}
Where $C$ is the measured conversion factor.  $C$ is between $7 - 12$
for most sources.  This method yields a flux of $2.9 \times 10^{-8}
\mathrm{ergs}\ \mathrm{cm^{-2}\ s^{-1}}$ for the Crab Nebula in the $2
- 10$ keV band, which is within $5\%$ of the accepted value of $2.8
\times 10^{-8} \mathrm{ergs}\ \mathrm{cm^{-2}\ s^{-1}}$.  Based on this result, we believe the 
value of $C$ to be correct in our observations at the $10\%$ level.  
Utilizing this conversion and assuming isotropic radiation we find:
\begin{equation}
\label{eq:lum}
	\log \left(\frac{L}{L_{\mathrm{Edd}}}\right) = \log \left(\frac{(\mathrm{USA\ rate})( C 
\times 10^{-12})(4\pi d^2)}{1.3\times 10^{38}M}\right)
\end{equation}
where $d$ is the distance to the source in cm and $M$ is the mass of the compact
object in $\Msolar$.  For the RXTE BHC data, a conversion factor was calculated to match the 
count rate per PCU in the PCA standard one data to the count rate seen by USA in the same 
source for an observation during the same day.  The average count rate per observation was 
used to calculate this conversion.  This conversion was calculated for each source to account 
for different detector response between the two experiments.  After the RXTE data was 
converted to USA rate then the  $\log \left(\frac{L}{L_{\mathrm{Edd}}}\right)$ was calculated 
following the method above. 

\begin{figure}
\begin{center}
\plotone{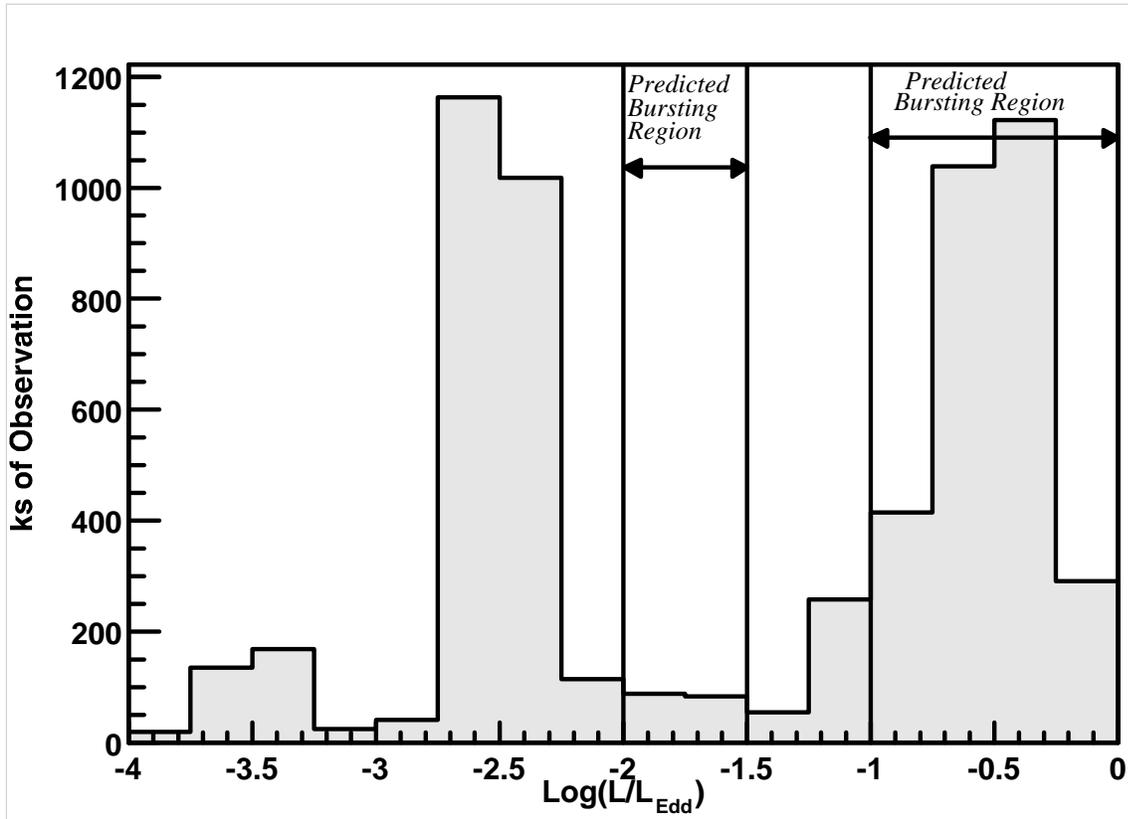}
\caption{Distribution of the $\log \left(\frac{L}{L_{\mathrm{Edd}}}\right)$ for our BHC 
observations.  The regions labeled ``Predicted Bursting Region'' are where \citet{nh02} 
find unstable burning that should lead to type I X-ray bursts if a surface exists on a 
$10\Msolar$ object.}
\label{fig:histo1}
\end{center}
\end{figure}

\begin{figure}
\begin{center}
\plotone{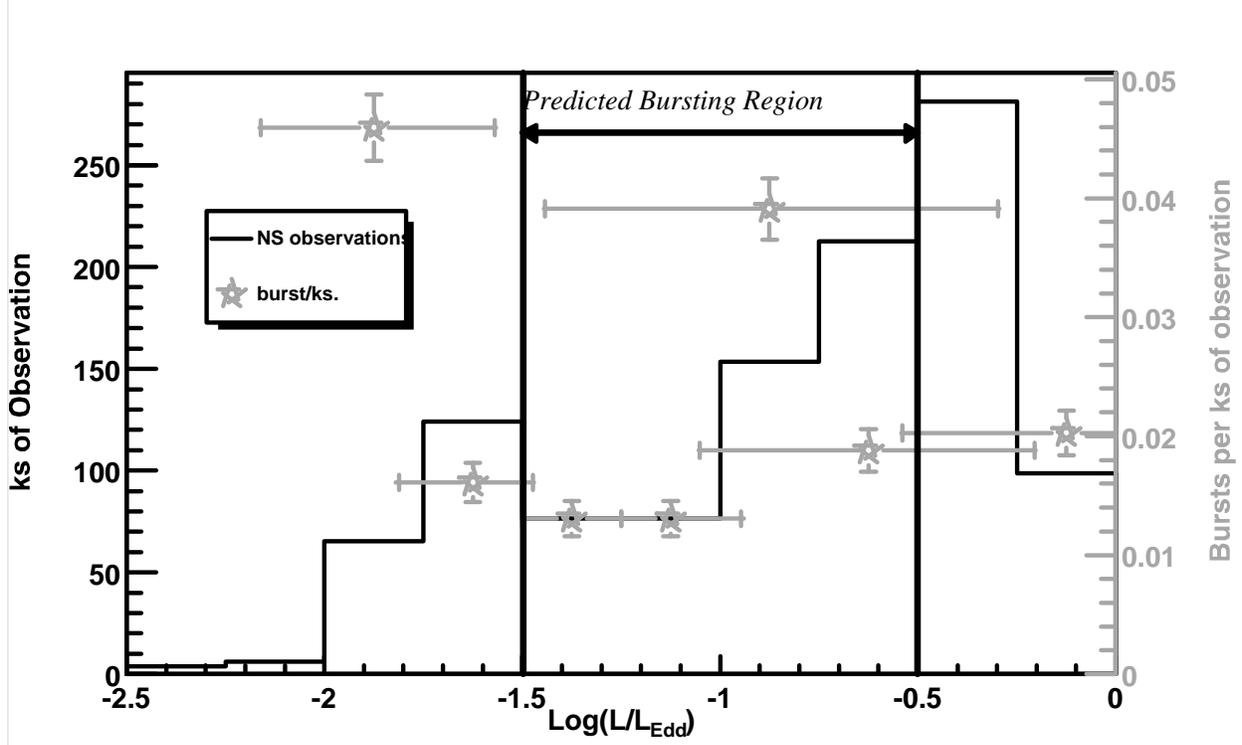}
\caption{
Distribution of the $\log \left(\frac{L}{L_{\mathrm{Edd}}}\right)$ for
our neutron star observations.  The solid histogram contains all the
observations of the seventeen neutron stars that we analyzed in
this paper.  The region labeled ``Predicted Bursting Region'' is where
\citet{nh02} find unstable burning that should lead to type I X-ray bursts.  
The gray stars and the right ordinate show where the nineteen bursts detected by USA were 
observed, and show a bursting rate in that particular bin.  The error bars reflect the 
uncertainties in the luminosity of the source(s) that had the burst(s) that went into the 
particular luminosity bin.}
\label{fig:histo2}
\end{center}
\end{figure}

Figures \ref{fig:histo1} and \ref{fig:histo2} give the resulting values of
$\log \left(\frac{L}{L_{\mathrm{Edd}}}\right)$ for the BHC data and
neutron star data analyzed in this paper.  Figure \ref{fig:histo2}
also shows the distribution of where we observed bursts in the data.\\

\subsubsection{Luminosity Uncertainties}

In order to estimate the uncertainty in
the amount of data that falls within the bursting region we use 
eq. \ref{eq:lum} to calculate an uncertainty in $\log
\left(\frac{L}{L_{\mathrm{Edd}}}\right)$ for each source and use this
uncertainty with Monte Carlo techniques .  The
distance and the mass contribute the largest systematic uncertainties
to $\log \left(\frac{L}{L_{\mathrm{Edd}}}\right)$.  The distances and
masses used and the errors are given in Table
\ref{tab:sources}. Uncertainties in $L$ are estimated using error
estimates on the mass $\sim 20\%$, if a definite uncertainty is not
quoted in Table \ref{tab:sources}, and the distance $\sim 20\%$, if a
definite uncertainty is not quoted in Table \ref{tab:sources}.  We
assume an uncertainty in $C$ of $10\%$, based on estimates of the Crab
luminosity, and assume the uncertainty in USA rate is negligible for all 
sources except GX 354$-$0 and the Rapid Burster.  For these two sources the 
counting rate of one was contaminated with counts from the other as 
discussed in \S \ref{sec:ns}.  We were able to deconvolve the two sources since
we know the total USA counting rate while pointed at each source and we know the 
factor of contamination from the source off axis.  For observations
where one of these two sources was directly observed and the other was not directly 
observed within a few days, the RXTE ASM counting rate was used to estimate the rate
of the off axis source.  To do this, the ASM rate was converted to USA rate
 using 72 ASM counts s$^{-1}$ as one Crab, and 3500 USA counts s$^{-1}$
as one Crab.  Therefore, for the Rapid Burster and GX 354$-$0, an error on the USA rate
of 20\% was used to calculate the errors on luminosity.  This accounts for changes in 
counting rate of the off axis source between observations, and/or errors in the ASM counting
rate.
For each observation, a Gaussian deviate was calculated with the same mean
as the measured value of $\log
\left(\frac{L}{L_{\mathrm{Edd}}}\right)$ and a $\ \sigma$ equal to the
uncertainty value calculated for that source.  Then the fraction of
this simulated data in the bursting region was measured.  We performed $10,000$
iterations in this manner and calculated the distribution of the
fraction of data in the bursting region.  We found that the
distribution of the amount of data within the bursting region was
rather tight, even though the individual errors on $\log
\left(\frac{L}{L_{\mathrm{Edd}}}\right)$ were large.  We found that, to
a confidence level of $99.5\%$, $25\%$ of our BHC data and $42\%$ of
our neutron star data fall within the region where one would expect
bursts according to \citet{nh02}.  Our analysis assumed that each of the sources
analyzed were accreting Hydrogen rich material.  If the accreted material has a large
fraction of Helium, as do some bursters, then this will increase the value of
 ${L_{\mathrm{Edd}}}$.\\

\subsection{BHC Surface Limit}

\citet{nh02}, calculated what would occur in a $10 \Msolar$ 
object if it were assumed to have a surface similar to that of a
neutron star, and showed that the rate of bursts should be comparable to
the rate in neutron stars.  However, they find different regions of
$L$ where bursts should occur.  For a $10 \Msolar$ object with a
surface and a base temperature of $10^7$ K, the regions where they
expect to see bursts are approximately: $-2 \leq \log
\left(\frac{L}{L_{\mathrm{Edd}}}\right) \lesssim -1.5$ and $-1
\lesssim \log \left(\frac{L}{L_{\mathrm{Edd}}}\right) \lesssim 0$.
For the BHC data analyzed, at least $25\%$ (1512 ks) of the data
fall within this range to a confidence level of $99.5\%$ using Monte
Carlo estimates of uncertainty described above.  Figure
\ref{fig:histo1} shows the distribution of the observations analyzed
in this paper in the units $\log \left(\frac{L}{L_{\mathrm{Edd}}}\right)$.

Considering only the neutron star data that fall in the luminosity
range corresponding to unstable nuclear burning, we find that the 19
bursts occurred in 464 ks of data.  This 464 ks value is the $99.5\%$
confidence level on the amount of data within the bursting region
based on the Monte Carlo techniques described above, thus accounting
for uncertainties in the value of $\log
\left(\frac{L}{L_{\mathrm{Edd}}}\right)$.

Using these data, we calculate the rate of bursting of neutron stars
in the unstable region $\lambda_{\mathrm{NS}} = 4.1\times10^{-5}$
bursts s$^{-1}$, $R_{NS} = 24.4$ ks or $\sim 7$ hours.  
Taking the \citet{nh02} prediction for the rate of bursting in a
BHC having a surface should be similar to that of a neutron star
provided both are within regions of unstable $L$, we can calculate the
probability of seeing no burst in our BHC data.  From
eq. \ref{eq:poisson}, the probability of observing zero bursts in
BHCs, if they exist with a rate similar to the rate observed in
neutron stars, is $e^{-\lambda_{\mathrm{NS}} T} = 1
\times 10^{-27}$, where $T$, the amount of BHC data residing in 
the unstable region, is 1512 ks.  Analyzing the data this way suggests
that BHC do not have a surface as described by
\citet{nh02} to a confidence level of $> 10 \ \sigma$.  The limits and measured 
X-ray bursting rates can be found in Table \ref{tab:limits}.  The columns
are the bursting rate or limit calculated using all data analyzed, and then
the bursting rate or limit calculated using only the data whose luminosity
is within the bursting range calculated by NH02.  This is the first such quantitative 
observational limit placed that quantifies what the lack of bursts means to the BHC theories.

One should point out that this is not definitive proof of the existence of black holes.  Other 
authors have pointed out that there are other states of matter that may exist in these massive 
compact objects that do not show X-ray bursts but also do not possess event horizons.  See 
\citet{akl02} for several arguments along these lines.  The authors in \citet{akl02} outline 
arguments that the accreted material could immediately be converted to some exotic form that 
would not show X-ray bursts \citep{arw98,rss+98}.  In addition, the argument is made that 
gravastars, if they exist, \citep{mm02} would be observational indistinguishable from black 
holes even though a gravastar exists without an event horizon or a singularity.    

\subsection{Neutron Star Observations Compared to Theory}

In order to test the validity of the \citet{nh02} predictions, we
compare the occurrence of the type I X-ray bursts in neutron stars to
the predicted occurrence calculated in
\citet{nh02}.  Specifically, after accounting for
errors in measuring $\frac{L}{L_\mathrm{Edd}}$, we see no evidence
that any of the observed neutron star bursts occurred when the source
was not in the region of unstable burning where bursts would be
expected.  We investigated the neutron stars in which bursts were not
observed, to determine if we expected to see bursts based on the
theory put forth in \citet{nh02}.  Of the neutron stars
analyzed where we did not see bursts, we acquired 60 ks of
data on GX 349+2
when it was in the region of
luminosity where one would expect bursts.  This is the largest amount
of individual neutron star data in the bursting region in which we did
not detect any bursts.  Again, using Poisson statistics and assuming a
bursting rate of $\lambda_{\mathrm{NS}} = 4.1\times10^{-5}$ bursts s$^{-1}$
calculated above, we find that we have a $\sim 10\%$ chance of not seeing a burst
in this source.  Therefore, there is some reason to
believe there is room to examine the boundaries of the bursting region predicting where the 
bursts
will occur.  However, it is worth noting that the USA data of X-ray binaries 
Scorpius X-1 and GX 349+2 are in the bursting region $43\%$ and $74\%$ of the time 
respectively (99.5\% confidence level).  In fact, one could use the fact that these are 
neutron stars that
have never been observed to burst to place limits on the distances to these
sources.  Assuming a mass of $1.4\Msolar$, GX 349+2 must not reside between
$5.9^{+0.6}_{-0.5}$ -- $9.7^{+1.0}_{-0.8}$ kpc
and Scorpius X-1 must not reside between$2.5^{+0.6}_{-0.4}$ -- $4.1^{+1.1}_{-0.6}$ kpc.
The errors are based on $1 \ \sigma$ deviations in the observed rate from these sources.  
Previous distance estimates for GX 349+2 place it around the galactic
center~5--8.5~kpc~\citep{cs97,cp91}, and estimates place Scorpius~X-1 
at~$2.8\pm 0.3$~kpc~\citep{gfb99}.  Both of these estimates place these sources at the 
edge of our excluded region.
The other two neutron star 
systems that we analyzed and have never been observed to burst, GX 340+0 and GX $5-1$
only spend $< 15\%$ and $< 2\%$ of their time in the bursting region.  Therefore,
it is likely that these sources are neutron stars that rarely, if ever, burst.

\section{Conclusions}
\label{sec:conclusions}

We present the burst catalog of the USA experiment X-ray binary data,
and calculate limits on the rate of X-ray bursts in BHCs.  We detected
nineteen type I X-ray bursts in seven neutron stars.  From our neutron star
data we conclude that the average time between bursts in neutron stars
is 59.1 ks, $\lambda_{NS} = 1.69\times10^{-5}$ bursts s$^{-1}$.  The
BHC data scanned showed no evidence for bursts.  Therefore, we place a
$95\%$ confidence limit on the burst rate in BHCs to be $\leq
4.9\times10^{-7}$ bursts s$^{-1}$ or $R \geq 470$ ks between bursts,
based on USA and RXTE BHC data.  The value of this upper limit is that
it is at a level that is not easily pushed lower.  This is because
most BHCs are transients that, with some exceptions, last less than a
few Megaseconds.  The few BHCs that are relatively steady (primarily Cygnus X-1
but arguably GX 339$-$4 and GRS 1915$-$105) do not have more than a few
tens of Megaseconds of observation.  Thus, $10^{-6} - 10^{-7}$ bursts s$^{-1}$
is a rough order of magnitude for the attainable limit.  A sensitive
all sky monitor might eventually reach a slightly lower value.\\

Applying the theoretical framework of \citet{nh02}, who
claim that if BHCs have surfaces they will burst as prolifically as
neutron stars if both are in a regime of unstable nuclear burning, we
find that the BHCs in this regime do not burst with the same or higher
rate as neutron stars to a confidence level of $> 10 \ \sigma$ based on our limits.
Therefore, these observations, analyzed according to the theory proposed by
\citet{nh02} leads us to believe that BHCs do not have a surface, as described in NH02, to a 
very high confidence level.

\acknowledgements
Work at SLAC was supported by Department of Energy contract DE-AC
03-76-SFO0515.  Basic research in X-ray astronomy at the Naval
Research Laboratory is supported by ONR/NRL.  This work was also
supported by the NASA Applied Information Systems Research Program.  
The authors would also like to thank Dr. Erik Kullkers, Dr. Lev Titarchuk,
Dr. Jean-Pierre Lasota and the anonymous referee for useful comments.

\bibliography{journapj,usarefs}

\begin{thebibliography}{}

\bibitem[\protect\citeauthoryear{{Abramowicz}, {Klu{\' z}niak}, \&
  {Lasota}}{{Abramowicz} et~al.}{2002}]{akl02}
{Abramowicz}, M.~A., {Klu{\' z}niak}, W.,  \& {Lasota}, J.-P. 2002, \aap, 396,
  L31

\bibitem[\protect\citeauthoryear{{Alford}, {Rajagopal}, \& {Wilczek}}{{Alford}
  et~al.}{1998}]{arw98}
{Alford}, M., {Rajagopal}, K.,  \& {Wilczek}, F. 1998, Physics Letters B, 422,
  247

\bibitem[\protect\citeauthoryear{{Augusteijn}, {Kuulkers}, \& {van
  Kerkwijk}}{{Augusteijn} et~al.}{2001}]{akv01}
{Augusteijn}, T., {Kuulkers}, E.,  \& {van Kerkwijk}, M.~H. 2001, \aap, 375,
  447

\bibitem[\protect\citeauthoryear{{Babak} \& {Grishchuk}}{{Babak} \&
  {Grishchuk}}{2002}]{bg02}
{Babak}, S.~V.,  \& {Grishchuk}, L.~P. 2002, gr-qc/0209006

\bibitem[\protect\citeauthoryear{{Bildsten}}{{Bildsten}}{2000}]{b00}
{Bildsten}, L. 2000, in Rossi2000: Astrophysics with the Rossi X-ray Timing
  Explorer. March 22-24, 2000 at NASA's Goddard Space Flight Center, Greenbelt,
  MD USA, p.E65

\bibitem[\protect\citeauthoryear{{Brandt} et~al.}{{Brandt}
  et~al.}{1992}]{bcl+92}
{Brandt}, S., {Castro-Tirado}, A.~J., {Lund}, N., {Dremin}, V., {Lapshov}, I.,
  \& {Syunyaev}, R. 1992, \aap, 262, L15

\bibitem[\protect\citeauthoryear{{Christian} \& {Swank}}{{Christian} \&
  {Swank}}{1997}]{cs97}
{Christian}, D.~J.,  \& {Swank}, J.~H. 1997, \apjs, 109, 177

\bibitem[\protect\citeauthoryear{{Cooke} \& {Ponman}}{{Cooke} \&
  {Ponman}}{1991}]{cp91}
{Cooke}, B.~A.,  \& {Ponman}, T.~J. 1991, \aap, 244, 358

\bibitem[\protect\citeauthoryear{{Cumming} \& {Bildsten}}{{Cumming} \&
  {Bildsten}}{2001}]{cb01}
{Cumming}, A.,  \& {Bildsten}, L. 2001, \apjl, 559, L127

\bibitem[\protect\citeauthoryear{{Dorman} \& {Arnaud}}{{Dorman} \&
  {Arnaud}}{2001}]{da01}
{Dorman}, B.,  \& {Arnaud}, K.~A. 2001, in ASP Conf. Ser. 238: Astronomical
  Data Analysis Software and Systems X, Vol.~10, 415

\bibitem[\protect\citeauthoryear{{Filippenko} \& {Chornock}}{{Filippenko} \&
  {Chornock}}{2001}]{fc01}
{Filippenko}, A.~V.,  \& {Chornock}, R. 2001, \iaucirc, 7644, 2

\bibitem[\protect\citeauthoryear{Fox}{Fox}{2003}]{fox03}
Fox, D.~D. 2003, Personal Communication

\bibitem[\protect\citeauthoryear{{Fox} et~al.}{{Fox} et~al.}{2001}]{flr+01}
{Fox}, D.~W., et~al. 2001, \mnras, 321, 776

\bibitem[\protect\citeauthoryear{{Galloway} et~al.}{{Galloway}
  et~al.}{2003}]{gpcm03}
{Galloway}, D.~K., {Psaltis}, D., {Chakrabarty}, D.,  \& {Muno}, M. 2003, \apj,
  in press, astro-ph/0208464

\bibitem[\protect\citeauthoryear{{Geldzahler}, {Fomalont}, \&
  {Bradshaw}}{{Geldzahler} et~al.}{1999}]{gfb99}
{Geldzahler}, B.~J., {Fomalont}, E.~B.,  \& {Bradshaw}, C.~F. 1999, Bulletin of
  the American Astronomical Society, 31, 897

\bibitem[\protect\citeauthoryear{{Greiner}, {Cuby}, \& {McCaughrean}}{{Greiner}
  et~al.}{2001}]{gcm01}
{Greiner}, J., {Cuby}, J.~G.,  \& {McCaughrean}, M.~J. 2001, \nat, 414, 522

\bibitem[\protect\citeauthoryear{{Guerriero} et~al.}{{Guerriero}
  et~al.}{1997}]{gfl+97}
{Guerriero}, R., et~al. 1997, American Astronomical Society Meeting, 191, 0

\bibitem[\protect\citeauthoryear{{Guerriero} et~al.}{{Guerriero}
  et~al.}{1999}]{gfk+99}
{Guerriero}, R., et~al. 1999, \mnras, 307, 179

\bibitem[\protect\citeauthoryear{{Hansen} \& {van Horn}}{{Hansen} \& {van
  Horn}}{1975}]{hv75}
{Hansen}, C.~J.,  \& {van Horn}, H.~M. 1975, \apj, 195, 735

\bibitem[\protect\citeauthoryear{{Herrero} et~al.}{{Herrero}
  et~al.}{1995}]{hkg+95}
{Herrero}, A., {Kudritzki}, R.~P., {Gabler}, R., {Vilchez}, J.~M.,  \&
  {Gabler}, A. 1995, \aap, 297, 556

\bibitem[\protect\citeauthoryear{{Iaria} et~al.}{{Iaria} et~al.}{2001}]{idbr01}
{Iaria}, R., {Di Salvo}, T., {Burderi}, L.,  \& {Robba}, N.~R. 2001, \apj, 548,
  883

\bibitem[\protect\citeauthoryear{Jahoda et~al.}{Jahoda et~al.}{1996}]{jsg+96}
Jahoda, K., Swank, J.~H., Giles, A.~B., Stark, M.~J., Strohmayer, T., Zhang,
  W.,  \& Morgan, E.~H. 1996, in Proc. SPIE, ed. O.~H. Siegmund \& M.~A.
  Gummin, Vol. 2808, 59

\bibitem[\protect\citeauthoryear{{Knight}}{{Knight}}{2003}]{kni03}
{Knight}, J. 2003, \nat, 422, 554

\bibitem[\protect\citeauthoryear{Kuulkers}{Kuulkers}{2003}]{kuu03}
Kuulkers, D.~E. 2003, Personal Communication

\bibitem[\protect\citeauthoryear{{Kuulkers} et~al.}{{Kuulkers}
  et~al.}{2002}]{kiv+02}
{Kuulkers}, E., et~al. 2002, \aap, 382, 503

\bibitem[\protect\citeauthoryear{{Kuulkers} \& {van der Klis}}{{Kuulkers} \&
  {van der Klis}}{2000}]{kv00}
{Kuulkers}, E.,  \& {van der Klis}, M. 2000, \aap, 356, L45

\bibitem[\protect\citeauthoryear{{Lewin}, {van Paradijs}, \& {Taam}}{{Lewin}
  et~al.}{1993}]{lvt93}
{Lewin}, W.~H.~G., {van Paradijs}, J.,  \& {Taam}, R.~E. 1993, Space Science
  Reviews, 62

\bibitem[\protect\citeauthoryear{{Lewin}, {van Paradijs}, \& {Taam}}{{Lewin}
  et~al.}{1995}]{lvt95}
{Lewin}, W. H.~G., {van Paradijs}, J.,  \& {Taam}, R.~E. 1995, in X-ray
  Binaries, ed. W.~H.~G. Lewin, J.~van Paradijs, \& E.~P.~J. van~den Heuvel
  (Cambridge University)

\bibitem[\protect\citeauthoryear{{Marshall} et~al.}{{Marshall}
  et~al.}{2001}]{mrf+01}
{Marshall}, H.~L., et~al. 2001, \aj, 122, 21

\bibitem[\protect\citeauthoryear{{McClintock} et~al.}{{McClintock}
  et~al.}{2001}]{mhg+01}
{McClintock}, J.~E., et~al. 2001, \apj, 555, 477

\bibitem[\protect\citeauthoryear{{Meyer-Hofmeister} \&
  {Meyer}}{{Meyer-Hofmeister} \& {Meyer}}{2001}]{mm01b}
{Meyer-Hofmeister}, E.,  \& {Meyer}, F. 2001, \aap, 372, 508

\bibitem[\protect\citeauthoryear{{Mignani} et~al.}{{Mignani}
  et~al.}{2002}]{mdcm02}
{Mignani}, R.~P., {De Luca}, A., {Caraveo}, P.~A.,  \& {Mirabel}, I.~F. 2002,
  \aap, 386, 487

\bibitem[\protect\citeauthoryear{{Mottola} \& {Mazur}}{{Mottola} \&
  {Mazur}}{2002}]{mm02}
{Mottola}, E.,  \& {Mazur}, P.~O. 2002, in American Physical Society, April
  Meeting, Jointly Sponsored with the High Energy Astrophysics Division (HEAD)
  of the American Astronomical Society April 20 - 23, 2002 Albuquerque
  Convention Center Albuquerque, New Mexico Meeting ID: APR02, abstract
  \#I12.011, 12011

\bibitem[\protect\citeauthoryear{{Narayan} \& {Heyl}}{{Narayan} \&
  {Heyl}}{2002}]{nh02}
{Narayan}, R.,  \& {Heyl}, J.~S. 2002, \apjl, 574, L139

\bibitem[\protect\citeauthoryear{{O'Neill} et~al.}{{O'Neill}
  et~al.}{2002}]{oksv02}
{O'Neill}, P.~M., {Kuulkers}, E., {Sood}, R.~K.,  \& {van der Klis}, M. 2002,
  \mnras, 336, 217

\bibitem[\protect\citeauthoryear{{Orosz} et~al.}{{Orosz} et~al.}{2002}]{ogv+02}
{Orosz}, J.~A., et~al. 2002, \apj, 568, 845

\bibitem[\protect\citeauthoryear{{Orosz} \& {Kuulkers}}{{Orosz} \&
  {Kuulkers}}{1999}]{ok99}
{Orosz}, J.~A.,  \& {Kuulkers}, E. 1999, \mnras, 305, 132

\bibitem[\protect\citeauthoryear{{Paczynski}}{{Paczynski}}{1983}]{p83}
{Paczynski}, B. 1983, \apj, 264, 282

\bibitem[\protect\citeauthoryear{{Peebles}}{{Peebles}}{2002}]{pee02}
{Peebles}, P. J.~E. 2002, astro-ph/0209403

\bibitem[\protect\citeauthoryear{{Rapp} et~al.}{{Rapp} et~al.}{1998}]{rss+98}
{Rapp}, R., {Sch{\" a}fer}, T., {Shuryak}, E.,  \& {Velkovsky}, M. 1998,
  Physical Review Letters, 81, 53

\bibitem[\protect\citeauthoryear{Ray et~al.}{Ray et~al.}{2001}]{rwf+01}
Ray, P.~S., et~al. 2001, in AIP Conference Proceedings, Vol. 599, X-ray
  Astronomy 1999 -- Stellar Endpoints, AGN, and the Diffuse Background, ed.
  G.~Malaguti, G.~Palumbo, \& N.~White (Melville, New York: AIP), 336

\bibitem[\protect\citeauthoryear{{Rutledge} et~al.}{{Rutledge}
  et~al.}{2000}]{rbb+00}
{Rutledge}, R.~E., {Bildsten}, L., {Brown}, E.~F., {Pavlov}, G.~G.,  \&
  {Zavlin}, V.~E. 2000, \apj, 529, 985

\bibitem[\protect\citeauthoryear{{Rutledge} et~al.}{{Rutledge}
  et~al.}{2001}]{rbb+01}
{Rutledge}, R.~E., {Bildsten}, L., {Brown}, E.~F., {Pavlov}, G.~G.,  \&
  {Zavlin}, V.~E. 2001, \apj, 559, 1054

\bibitem[\protect\citeauthoryear{{Saz Parkinson}}{{Saz
  Parkinson}}{2003}]{saz03}
{Saz Parkinson}, P.~M. 2003, Ph.D. thesis, Stanford Universtiy, In Press

\bibitem[\protect\citeauthoryear{{Schmutz}, {Geballe}, \& {Schild}}{{Schmutz}
  et~al.}{1996}]{sgs96}
{Schmutz}, W., {Geballe}, T.~R.,  \& {Schild}, H. 1996, \aap, 311, L25

\bibitem[\protect\citeauthoryear{{Seon} et~al.}{{Seon} et~al.}{1997}]{smy+97}
{Seon}, K., {Min}, K., {Yoshida}, K., {Makino}, F., {Lewin}, W.~H.~G., {van der
  Klis}, M.,  \& {van Paradijs}, J. 1997, \apj, 479, 398

\bibitem[\protect\citeauthoryear{{Shahbaz} \& {Kuulkers}}{{Shahbaz} \&
  {Kuulkers}}{1998}]{sk98}
{Shahbaz}, T.,  \& {Kuulkers}, E. 1998, \mnras, 295, L1

\bibitem[\protect\citeauthoryear{{Singh} et~al.}{{Singh} et~al.}{2002}]{snp+02}
{Singh}, N.~S., {Naik}, S., {Paul}, B., {Agrawal}, P.~C., {Rao}, A.~R.,  \&
  {Singh}, K.~Y. 2002, \aap, 392, 161

\bibitem[\protect\citeauthoryear{Strohmayer \& Bildsten}{Strohmayer \&
  Bildsten}{2003}]{sb03}
Strohmayer, T.,  \& Bildsten, L. 2003, in Compact Stellar X-ray Sources, ed.
  W.~H.~G. Lewin \& M.~van~der Klis (Cambridge University Press), in press
  astro-ph/0301544

\bibitem[\protect\citeauthoryear{{Strohmayer} \& {Brown}}{{Strohmayer} \&
  {Brown}}{2002}]{sb02}
{Strohmayer}, T.~E.,  \& {Brown}, E.~F. 2002, \apj, 566, 1045

\bibitem[\protect\citeauthoryear{{Sunyaev} \& {Revnivtsev}}{{Sunyaev} \&
  {Revnivtsev}}{2000}]{sr00}
{Sunyaev}, R.,  \& {Revnivtsev}, M. 2000, \aap, 358, 617

\bibitem[\protect\citeauthoryear{{Titarchuk}, {Mastichiadis}, \&
  {Kylafis}}{{Titarchuk} et~al.}{1997}]{tmk97}
{Titarchuk}, L., {Mastichiadis}, A.,  \& {Kylafis}, N.~D. 1997, \apj, 487, 834

\bibitem[\protect\citeauthoryear{{Titarchuk} \& {Shaposhnikov}}{{Titarchuk} \&
  {Shaposhnikov}}{2002}]{ts02}
{Titarchuk}, L.,  \& {Shaposhnikov}, N. 2002, \apjl, 570, L25

\bibitem[\protect\citeauthoryear{{Ubertini} et~al.}{{Ubertini}
  et~al.}{1999}]{ubc+99}
{Ubertini}, P., {Bazzano}, A., {Cocchi}, M., {Natalucci}, L., {Heise}, J.,
  {Muller}, J.~M.,  \& {in 't Zand}, J.~J.~M. 1999, \apjl, 514, L27

\bibitem[\protect\citeauthoryear{{van der Klis}}{{van der Klis}}{2000}]{van00}
{van der Klis}, M. 2000, \araa, 38, 717

\bibitem[\protect\citeauthoryear{{van Paradijs} \& {White}}{{van Paradijs} \&
  {White}}{1995}]{vw95}
{van Paradijs}, J.,  \& {White}, N. 1995, \apjl, 447, L33

\bibitem[\protect\citeauthoryear{{Wagner} et~al.}{{Wagner}
  et~al.}{2001}]{wfs+01}
{Wagner}, R.~M., {Foltz}, C.~B., {Shahbaz}, T., {Casares}, J., {Charles},
  P.~A., {Starrfield}, S.~G.,  \& {Hewett}, P. 2001, \apj, 556, 42

\bibitem[\protect\citeauthoryear{{Wijnands} et~al.}{{Wijnands}
  et~al.}{2002}]{wmm+02}
{Wijnands}, R., {Muno}, M.~P., {Miller}, J.~M., {Franco}, L.~., {Strohmayer},
  T., {Galloway}, D.,  \& {Chakrabarty}, D. 2002, \apj, 566, 1060

\bibitem[\protect\citeauthoryear{{Zurita} et~al.}{{Zurita}
  et~al.}{2002}]{zsc+02}
{Zurita}, C., et~al. 2002, \mnras, 334, 999

\end{thebibliography}

\clearpage

\begin{deluxetable}{lcccll} 
\tabletypesize{\small}
\tablecaption{USA observing times of X-ray binaries searched for bursts}

\startdata
\textbf{Source}	& \textbf{Time}	& \textbf{Dist.} & \textbf{Mass} 
&\textbf{Class}\tablenotemark{c} & \textbf{Reference(s)}\\   
				& \textbf{(ks)}& \textbf{(kpc)} & \textbf{estimate} $\mathbf{(\Msolar)}$ &			 
& 	\\ \tableline\tableline \\[1 pt]
Cygnus X-1	& 449 - USA & 2 & $\sim 10.1 $ &BHC & {\citealt{hkg+95}}\\	
	&	1748 - RXTE	&\\ \tableline \\[-9 pt]
XTE J1118+480& 183 - USA&$1.9 \pm 0.4$&$6.0 \pm 0.36 $  &BHC & {\citealt{wfs+01}}\\	
			   &	119 - RXTE	  &	  &			&	&			  
{\citealt{mhg+01}}\\\tableline \\[-9 pt]
GRS 1915+105 & 163 - USA	&12.5&$14 \pm 4 $ & BHC & {\citealt{gcm01}}\\
			 & 2358 - RXTE	& \\	\tableline \\[-9 pt]
XTE J1859+226& 94 - USA	&11& $\geq 7.4 \pm 1.1 $  &BHC& {\citealt{fc01}}\\
			   &311 - RXTE  &	   & 	   	   	   				&   & {\citealt{zsc+02}}\\ 
\tableline \\[-9 pt]
XTE J1550$-$564& 50 - USA	& $3-6$& $9.7 - 11.6 $ &BHC & {\citealt{ogv+02}}\\ 
			   & 606 - RXTE & \\	\tableline \\[-9 pt]   
4U 1630$-$472& 46.2 - USA	&10 &??\tablenotemark{a}&BHC  & {\citealt{akv01}}\\	
   			   &  &	   & 	   	&   & {\citealt{mm01b}}\\ \tableline \\[-9 pt]
Cygnus X-3	& 37.0 - USA	&11.6 &$17^{+23}_{-10}\tablenotemark{d} $&BHC   & 
{\citealt{snp+02}}\\	
	   		& 335 - RXTE		&	  & 	  &	&{\citealt{sgs96}}\\ \tableline \\[-9 pt]
Cygnus X-2& 158.1 & $7.2 \pm 0.11$ &$1.4 \pm 0.6 $ &NS, b&{\citealt{ts02}}\\ 
	   	&		  &		 	 	   &	 	 			 &	  &{\citealt{ok99}}\\ \tableline 
\\[-9 pt]
Aquila X-1	& 93.2	&$4-6.5$ &$\sim 1.4 $\tablenotemark{b}&NS A, b   & {\citealt{rbb+01}}\\	
\tableline \\[-9 pt]
EXO 0748$-$676& 124.8	& 7.6 &$\sim 1.4 $\tablenotemark{b}&NS, b   & {\citealt{vw95}}\\	
			  		& 	&	   			&					&   & {\citealt{sk98}}\\  
\tableline \\[-9 pt]
Rapid Burster & 89.2 &8.6 &$\sim 1.4 $\tablenotemark{b}&NS, b   & {\citealt{mrf+01}}\\	
\tableline \\[-9 pt]
GX 354$-$0	& 81.1	& $4.4 - 6.2$&$\sim 1.4 $\tablenotemark{b}&NS A, b& {\citealt{gpcm03}}\\ 
\tableline \\[-9 pt]	
Circinus X-1	& 75.6	& $6.7\pm 1.2$&$1.4 - 3.0 $\tablenotemark{e}&NS, b & 
{\citealt{mdcm02}}\\
		 		& 		& 		  	  &		  							  &      
&{\citealt{saz03}}\\	\tableline \\[-9 pt]
4U 0614+09& 66.2  & 5& $\sim 1.4 $\tablenotemark{b}&NS, b &{\citealt{bcl+92}}\\ 
   		&		  &	 & 		 	 					&     &{\citealt{cs97}}\\\tableline \\[-9 
pt]
GX 349+2 &  81.6&5&$\sim 1.4 $\tablenotemark{b}&NS Z & {\citealt{oksv02}}\\  
(Scorpius X-2)	& & & 							& 	 & {\citealt{cs97}}\\ \tableline \\[-9 pt]
XB 1254$-$690& 51.7 & 12 &??\tablenotemark{f}&NS, b&{\citealt{idbr01}}\\ 
			 &	  &	   &	  	  					 &	   &{\citealt{cs97}}	\\ \tableline 
\\[-9 pt]
Scorpius X-1& 44.3 &$2.8 \pm 0.3$  &$1.4 \pm 0.6 $ &NS, Z&{\citealt{ts02}}\\
		 	 &	   &	 	 	   &	 	 			 &	& {\citealt{gfb99}}\\ \tableline 
\\[-9 pt] 
4U 1735$-$445  	& 38.1	& $9.2$&$\sim 1.4 $\tablenotemark{b}&NS A, b   & {\citealt{smy+97}}\\
   				& 		& 		&	  	  					&	       & 
{\citealt{vw95}}\\\tableline \\[-9 pt]	
X1636$-$536& 37.9 & $6.5\pm 0.2$ &$\sim 1.4 $\tablenotemark{b} &NS, b&{\citealt{vw95}}\\
			 &	  &				 &							   &	&{\citealt{cs97}}\\ 
\tableline \\[-9 pt]
GX 5$-$1  & 34.6 & 9 & $\sim 1.4 $\tablenotemark{b}&NS, Z&{\citealt{cs97}}	\\ \tableline 
\\[-9 pt]
GX 3+1	& 28.7	&4-6 &$\sim 1.4 $\tablenotemark{b}&NS A, b   & {\citealt{kv00}}\\
   		& 		&	 &&	  		  & {\citealt{cs97}}\\ \tableline \\[-9 pt]
GX 340+0& 28.5 & 9.5-11.0 & $\sim 1.4 $\tablenotemark{b}&NS, Z&{\citealt{cs97}}	\\ \tableline 
\\[-9 pt]
GX 17+2 & 29.4 & 7.5 & $\sim 1.4 $\tablenotemark{b}&NS, Z, b&{\citealt{cs97}}\\ \tableline
MXB 1659$-$298& 16.0	&$\sim 10$ &$\sim 1.4 $\tablenotemark{b}&NS A, b   & 
{\citealt{wmm+02}}\\ \tableline \\[-9 pt]	
\enddata

\tablenotetext{a}{There is no dynamical mass estimate for 4U 1630$-$472.  This source was not 
used in burst rate limit calculations.}
\tablenotetext{b}{This neutron star mass is unknown, a value of $1.4\Msolar$ was assumed.}
\tablenotetext{c}{Class of the source: BHC - Black Hole Candidate; NS - Neutron Star; A - 
Atoll source; Z - Z source; b - if bursts have been detected in the source in the past.}
\tablenotetext{d}{The mass function for this source is $2.3 \Msolar$, the nature of the 
compact object in this source is still unknown.  This source was not used in burst rate limit 
calculations.}
\tablenotetext{e}{Timing and spectral 
analysis of this source give evidence that Circinus X-1 is larger than $1.4 \Msolar$
 \citep{saz03}.  This source was not used in burst rate calculations} 
 \tablenotetext{f}{The nature of this source could not be identified independent of bursts, 
therefore this source was not used in burst rate calculations.}
\vspace*{+0.5cm}
\label{tab:sources} 
\end{deluxetable}

\begin{table*}
\begin{center}
\caption{Detected type I X-ray bursts of the USA experiment}
\begin{tabular}{lcl} 
\tableline \tableline
\textbf{Source} & \textbf{Time of Burst} & \textbf{Rate Change}\\
				& \textbf{Peak [MJD]}    & 	\textbf{[cts s$^{-1}$]}	\\ \tableline
Aquila X-1	  & 51856.15684 & $1000 \longrightarrow 5000$  \\ \tableline
EXO 0748$-$676 & 51614.54328  & $30 \longrightarrow 420$  \\ 
			 & 51648.97544  & $40 \longrightarrow 420$  \\ 
			 & 51648.41676  & $35 \longrightarrow 400$  \\ 
			 & 51669.44317  & $40 \longrightarrow 470$  \\ 
			 & 51687.50888  & $50 \longrightarrow 520$  \\ \tableline
Rapid Burster / & 51482.51368 & $300 \longrightarrow 2700$  \\ 
GX 354$-$0\tablenotemark{a}	  & 51487.23736 & $200 \longrightarrow 2100$  \\ 
			  & 51492.94726 & $150 \longrightarrow 2100$  \\
			  & 51861.23391 & $220 \longrightarrow 1100$  \\
			  & 51861.93667 & $120 \longrightarrow 1500$   \\ 
			  & 51491.82323 & $300 \longrightarrow 2000$  \\
   		 	  & 51743.31991 & $250 \longrightarrow 2600$  \\
   		      & 51812.96651 & $250 \longrightarrow 3000$  \\
   		      & 51816.00111 & $300 \longrightarrow 4000$  \\ 
   		      & 51817.41458 & $250 \longrightarrow 4400$  \\ \tableline
4U 1735$-$445 	  & 51379.11730 & $400 \longrightarrow 3400$  \\ \tableline
GX 3+1		  & 51396.12709 & $200 \longrightarrow 1000$  \\ \tableline
MXB 1659$-$298  & 51850.08140 & $180 \longrightarrow 400$  \\ \tableline
\end{tabular}
\tablenotetext{a}{
The Rapid Burster and GX 354$-$0 are both in the USA FOV at the same
time, therefore a burst in one source would be indistinguishable from
a burst in the other.}
\vspace*{+0.5cm}

\label{tab:bursts} 
\end{center} 
\end{table*}

\begin{deluxetable}{lccl} 
\tablecaption{X-ray Burst Rates and Limits from the USA \& RXTE Experiments}
\startdata
	& \textbf{All Data}	& \textbf{Data in Predicted Bursting Range}\\ 
	&  \textbf{[bursts s$^{-1}$]}& \textbf{[bursts s$^{-1}$]} \\\tableline\tableline \\[1 pt]
\textbf{Neutron Stars} & $1.69\times10^{-5}$	&  $4.1\times10^{-5}$\\ \tableline
\textbf{Black Hole Candidates}\tablenotemark{a}& $<4.9\times10^{-7}$&
 $<2.0\times10^{-6}$\\ \tableline
\enddata
\tablenotetext{a}{The 95\% confidence level upper limit on the bursting rate in the BHCs.} 
\vspace*{+0.5cm}
\label{tab:limits} 
\end{deluxetable}

\clearpage

\end{document}